\documentclass[aps,preprint,showpacs,amsmath,amssymb]{revtex4-1}
\usepackage{amssymb}
\usepackage{epsf}
\usepackage[english]{babel}
\usepackage{amsmath}

\begin{document}

\newcommand {\half} {\mbox{$\frac{1}{2}$}}

\title{Connection between effective-range expansion
and nuclear vertex constant or asymptotic normalization coefficient}
\author{R. Yarmukhamedov}
\email{rakhim@inp.uz}
\affiliation{Institute of Nuclear Physics, Uzbekistan Academy of
Sciences, 100214 Tashkent, Uzbekistan}
\author{D. Baye}
\email{dbaye@ulb.ac.be}
\affiliation{Physique Quantique, CP165/82 and
Physique Nucl\'eaire Th\'eorique et Physique Math\'ematique, CP 229,
Universit\'e Libre de Bruxelles (ULB), B 1050 Bruxelles, Belgium}
\date{\today}

\begin{abstract}
Explicit relations between the effective-range expansion and
the nuclear vertex constant or asymptotic normalization coefficient (ANC)
for the virtual decay $B\to A+a$ are derived for an  arbitrary orbital momentum
together with the corresponding location condition for the ($A+a$) bound-state energy.
They are valid both for the charged case and for the neutral case.
Combining these relations with the standard effective-range function
up to order six makes it possible to reduce to two the number of free
effective-range parameters if an ANC value is known from experiment.
Values for the scattering length, effective range, and form parameter
are determined in this way for the $^{16}$O+$p$, $\alpha$+$t$ and $\alpha$+$^3$He
collisions in partial waves where a bound state exists
by using available ANCs deduced from experiments.
The resulting effective-range expansions for these collisions are valid
up to energies larger 5 MeV.
\end{abstract}


\maketitle

\section{Introduction}

The effective-range expansion provides a model-independent description
of a low-energy phase shift in a given partial wave for the $Aa$ scattering
\cite{BJ1949,Bethe1949,PB1975}.
Nevertheless, a lot of ambiguity occurs in the determination of the coefficients
of this expansion.
Several sets of parameters can describe equally well the low-energy phase shifts.
One possibility for removing this ambiguity lies in a consistent description
of the low-energy scattering of two particles and of the fundamental characteristics
of a two-body ($A+a$) weakly bound  state composed of these two particles
\cite{Igam1997,Igam2010,Spar2009,Orlov2010}.
The essential quantities required for such a description are the
binding energy of this state and its nuclear vertex constant (NVC) 
or its asymptotic normalization coefficient (ANC) 
which determines the amplitude of the tail of the
bound-state wave function in the two-particle channel.
The introduction of these two experimental values into a phase-shift
analysis performed with the effective-range expansion could allow one
to reduce the ambiguity for the first coefficients of this expansion.
Moreover, the first coefficients obtained in such a way should be helpful
to test the validity of microscopic models \cite{KB2007} or to constrain
properties of two-body potentials.

In Refs.~\cite{Igam1997,Igam2010,Spar2009,Orlov2010},
different forms of the ANC for the $A+a \to B$ vertex
with two charged particles ($A$ and $a$) and an arbitrary orbital
momentum $l$ were derived for the standard effective-range function.
For example, two expressions are derived in Ref.~\cite{Spar2009} for the ANC,
one for the neutral case and another one for the charged case.

In the present work, the results of Ref.~\cite{Spar2009} are generalized
for an effective-range function which is valid both for the charged
and neutral cases.
Combining this expression with the bound-state condition on the effective-range
expansion and taking into account the additional information about ``experimental''
values of the ANC for $A+a \to B$ makes  it possible to reduce the number of
free parameters in the expansion to two in the effective-range function
restricted up to order six in the momentum $k$.

As applications, we consider the $^{16}{\rm O}+p$, $\alpha+t$ and $\alpha+^3$He
collisions for which ``experimental'' ANC-values for the virtual decays
$^{16}{\rm O}+p \to ^{17}$F, $\alpha+t \to ^7$Li, and $\alpha + {^3{\rm He}}\to ^7$Be,
respectively, are known.
The values of the effective-range expansion parameters obtained in such a way
should be reliable.

In Sec.~\ref{ere}, the explicit expression for the nuclear vertex constant (or the respective ANC)
for the virtual decay $B\to A+a$ with two charged particles $A$  and $a$ and an arbitrary
relative angular momentum are derived for the standard effective-range expansion.
Detailed expressions restricted to terms up to $k^6$ are established.
They are also valid for the neutral case.
The results of the application of these expressions to various concrete scatterings
of light nuclei, for which the corresponding ANCs are known, are presented in Sec.~\ref{app}.
A conclusion is given in Sec.~\ref{conc}.

\section {Effective-range expansion and nuclear vertex constant}
\label{ere}

Let us consider two particles $A$ and $a$ of charges
$Z_A$ and $Z_a$, respectively, with a reduced mass $\mu $.
Let $k$ be the relative momentum of particles $A$ and $a$.
The  center-of-mass energy is $E=k^2/2\mu$ and
the dimensionless Sommerfeld  parameter is $\eta =Z_AZ_a e^2\mu/k$.
Let us denote by $l$ ($j$) the orbital (total) angular momentum
of the relative motion of these particles, by $\delta_{lj}$
the Coulomb modified nuclear phase shift, and by
$\sigma_l = \arg\Gamma(l+1+i\eta)$
the Coulomb phase shift for the $Aa$ scattering.
Everywhere we use the unit $\hbar = 1$.

 The partial scattering matrix or $S$ matrix $S_{lj}$ in the presence of both a
Coulomb and a nuclear interaction is determined by \cite{New}
\begin{equation}
S_{lj}=e^{2i(\delta_{lj}+\sigma_l)}
= \frac{\Gamma(l+1+i\eta)}{\Gamma(l+1-i\eta)}\,\,\frac{{\rm
{cot}}\,\delta_{lj}+i}{{\rm {cot}}\,\delta_{lj}-i}\,.
\label{eq1}
\end{equation}
Since this $S$ matrix has a rather complicated analytical structure
in the complex $E$-plane, it is useful to introduce a function
with simpler analytical properties \cite{Ham1973,Haer1977},
\begin{equation}
F_{lj}(k^2) = \frac{e^{2i\delta_{lj}}-1}{2i}\,
\frac{l!^2e^{2i\sigma_l}e^{\pi\eta}}{k^{2l+1}\Gamma^2(l+1+i\eta)}.
\label{eq2}
\end{equation}
This function can be rewritten in the form
\begin{equation}
F_{lj}(k^2)=\frac{1}{k^{2l+1}C^{2}_0(\eta) D_l(\eta)({\rm
{cot}}\,\delta_{lj}-i)}
\label{eq3}
\end{equation}
with the definitions
\begin{equation}
C^{2}_0(\eta)=\frac{2\pi\eta}{e^{2\pi\eta}-1},\,\,\,\,
D_{l>0}(\eta)=\prod\limits_{n=1}^{l }\left(1+\eta^2/n^2\right).
\label{eq4}
\end{equation}
For the $s$ wave ($l=0$), the factor $D_0(\eta)$ in Eq.~(\ref{eq3}) is unity.
The function $D_l$ are related to the functions $w_l$ used in Ref.~\cite{Spar2009}
by $D_l = \eta^{2l} w_l /l!^2$.
While $w_l$ does not require a special treatment of $l=0$ and is very convenient
to study the limit $E \rightarrow 0$ \cite{BB2000}, $D_l$ does not require a separate
treatment of the neutral case $\eta = 0$.

The effective-range function is defined as
\begin{eqnarray}
K_{lj}(k^2)=\frac{1}{F_{lj}(k^2)}+2\eta D_l(\eta)H(\eta)k^{2l+1}
\label{eq5} \\
= k^{2l+1}D_l(\eta)[C^{2}_0(\eta){\rm
{cot}}\,\delta_{lj}+2\eta h(\eta)].
\label{eq6}
\end{eqnarray}
Here
\begin{equation}
H(\eta)=\psi(i\eta)-\ln(i\eta)+1/2i\eta=h(\eta)+\frac{i}{2\eta}C^2_0(\eta),
\label{eq7}
\end{equation}
where $\psi$ is the digamma function \cite{Abr1965},
\begin{equation}
h(\eta)=-\gamma+\eta^2\sum\limits_{n=1}^{\infty}\frac{1}{n(n^2+\eta^2)}-\ln\eta,
\end{equation}
and $\gamma=0.57721\dots$ is Euler's constant.
When $\eta$ is real, $h(\eta)$ is the real part of $H(\eta)$.

The Coulomb-nuclear partial elastic-scattering amplitudes are defined as
\begin{equation}
M_{lj}(E)=\frac{i\pi}{\mu k}\,e^{2i\sigma_l}\left(e^{2i\delta_{lj}}-1\right).
\label{eq8}
\end{equation}
They are related to the $F_{lj}(k^2)$ function by Eq.~(\ref{eq2}) and thus to
the effective-range function $K_{lj}(k^2)$ in Eq.~(\ref{eq5}) by
\begin{equation}
M_{lj}(E)=-\frac{2\pi}{\mu}\, \frac{k^{2l}\Gamma^2(l+1+i\eta)e^{-\pi\eta}}
{l!^2[K_{lj}(k^2)-2\eta D_l(\eta)H(\eta)k^{2l+1}]}.
\label{eq9}
\end{equation}

For negative energies $E=-\varepsilon$, where $\varepsilon>0$ is the binding
energy of the bound state of nucleus $B$ in the ($A+a$)-channel,
bound states correspond to poles of the scattering partial $S_{lj}$ matrix
(or the partial amplitude $M_{lj}$) on the positive imaginary $k$ axis
(or the negative $E$ axis) \cite{Haer1977}.
Let $k=i\kappa$  be the location of such a pole,
where $\kappa=\sqrt{2\mu\varepsilon}$.
According to Ref.~\cite{Iw1984}, it follows from Eqs.~(\ref{eq8})
and (\ref{eq3}) that this bound state corresponds to a zero of
$F_{lj}^{-1}(-\kappa^2)$.
Hence, from the denominator in Eq.(\ref{eq9}), one obtains
\begin{equation}
(-1)^l\kappa^{2l+1}=-\frac{K_{lj}(-\kappa^2)}{J_l(\eta_B)},
\label{eq10}
\end{equation}
where $\eta_B=Z_AZ_a e^2\mu/\kappa$ is the Sommerfeld parameter for
the ($A+a$) bound state and the real function $J_l(\eta_B)$ is defined by
\begin{equation}
J_l(\eta_B) = -2\eta_B D_l(-i\eta_B) H(-i\eta_B)
\end{equation}
with
\begin{equation}
H(-i\eta_B) = {\rm Re\,}h(-i\eta_B) - \half \pi \cot\pi\eta_B.
\end{equation}
Equation (\ref{eq10}) is the pole-location condition.

According  to Ref.~\cite{Blokh77}, the residue of the partial amplitude
$M_{lj}(E)$ at the pole $E=-\varepsilon\,\,\,(k=i\kappa)$ is
expressed in terms of the NVC $G_{lj}$ for the virtual decay
$B\to A+a$ as
\begin{equation}
{\rm res\,} M_{lj}(E)
= \lim_{E\to -\varepsilon} (E+\varepsilon) M_{lj}(E)
= G_{lj}^2.
\label{eq11}
\end{equation}
Alternatively, the NVC can be obtained from the scattering matrix $S_{lj}$
through the relation
\begin{equation}
{\rm res\,} S_{lj}(E) = \frac{\mu\kappa}{\pi} {\rm res\,} M_{lj}(E).
\label{eq11a}
\end{equation}
Combining Eqs.~(\ref{eq1}), (\ref{eq8}), (\ref{eq9}), and (\ref{eq11}), one obtains
\begin{equation}
G_{lj}=i^{l+\eta_B}\sqrt{\frac{2\pi}{\mu}}\frac{\kappa^l\Gamma(l+1+\eta_B)}{l!}\left[-\frac{dF_{lj}^{-1}(k^2)
}{dE}\Big|_{E=-\varepsilon}\right]^{-1/2}.
\label{eq12}
\end{equation}
Differentiating function $F_{lj}^{-1}(k^2)$ determined from Eqs.~(\ref{eq8}) and (\ref{eq9})
leads for $E=-\varepsilon$ ($k=i\kappa$) to the explicit expression for the NVC,
\begin{equation}
G_{lj}^2=(-1)^le^{i\pi\eta_B}\frac{2\pi}{\mu^2}\frac{\kappa^{2l+1}\Gamma^2(l+1+\eta_B)}{l!^2}
\left[(-1)^l\kappa^{2l}f_l(\eta_B)-\frac{\kappa}{\mu}\frac{dK_{lj}(k^2)
}{dE}\Big|_{E=-\varepsilon}\right]^{-1},
\label{eq13}
\end{equation}
where
\begin{equation}
f_l(\eta_B)= D_l(-i\eta_B) \left\{
\left(\frac{\pi\eta_B}{\sin\pi\eta_B}\right)^2 - 2\eta_B \left[\tilde
{h}(\eta_B)+2(1-\delta_{l0})H(-i\eta_B)
\left(l+ \sum_{n=1}^l \frac{\eta_B^2}{n^2-\eta_B^2} \right) \right] \right\},
\label{eq14}
\end{equation}
and
\begin{equation}
\tilde
{h}(\eta_B)= -\left[\eta \frac{dh}{d\eta} \right]_{\eta=-i\eta_B}
= 1 + 2\eta_B^2\sum\limits_{n=1}^{\infty}\frac{1}{n(n^2-\eta_B^2)}
+2\eta_B^4\sum\limits_{n=1}^{\infty}\frac{1}{n(n^2-\eta_B^2)^2}.
\label{eq15}
\end{equation}

Equations (\ref{eq10}) and (\ref{eq13}) are quite general. We now
particularize them to low binding and scattering energies by using
Taylor expansions of the effective-range function $K_{lj}(k^2)$
for $k^2\to$ 0 and we then keep terms up to $k^6$ in this expansion,
\begin{equation}
K_{lj}(k^2)\approx
-\frac{1}{a_{lj}}+\frac{r_{lj}}{2}k^2-P_{lj}r_{lj}^3 k^4+Q_{lj}k^6,
\label{eq16}
\end{equation}
where the scattering length $a_{lj}$ is  in fm$^{2l+1}$,
the effective range $r_{lj}$ is in fm$^{-2l+1}$,
the form parameter $P_{lj}$ is in fm$^{4l}$,
and the sixth-order coefficient $Q_{lj}$ is in fm$^{-2l+5}$.
In this approximation, Eqs.~(\ref{eq10}) and (\ref{eq13})
can be reduced to the forms
\begin{equation}
(-1)^l\kappa^{2l+1}\approx \frac{1/a_{lj}+\half r_{lj}\kappa^2
+P_{lj}r_{lj}^3\kappa^4+Q_{lj}\kappa^6}{J_l(\eta_B)}
\label{eq19}
\end{equation}
and
\begin{equation}
G_{lj}^2 \approx (-1)^le^{i\pi\eta_B}\frac{2\pi}{\mu^2}\frac{\kappa^{2l+1}\Gamma^2(l+1+\eta_B)}
{l!^2 {\cal B}_l(\kappa,\eta_B;r_{lj},P_{lj},Q_{lj})},
\label{eq17}
\end{equation}
where
\begin{equation}
{\cal B}_l(\kappa,\eta_B;r_{lj},P_{lj},Q_{lj})=(-1)^l\kappa^{2l}f_l(\eta_B)
-r_{lj}\kappa-4P_{lj}r_{lj}^3\kappa^3-6Q_{lj}\kappa^5.
\label{eq18}
\end{equation}
It should be noted that, for the case $P_{lj}=Q_{lj}=0$,  equations (\ref{eq17}) and
(\ref{eq19}) coincide with expressions (31) of Ref.~\cite{Igam1997} and (25)
of Ref.~\cite{Igam2010}, respectively, if the comments made in Ref.~\cite{Igam2010}
are taken into account.
The normalization for the partial amplitude (\ref{eq17}) differs
from that chosen in Ref.~\cite{Igam1997} by a Coulomb-phase multiplicative
factor $e^{2i\sigma_l}$. The allowance of this factor in the corresponding expressions
of Ref.~\cite{Igam1997} results in the replacement of  the factor $K(\eta_B)$,
entering in the numerator of the right-hand side of expressions (31) and (34)
of Ref.~\cite{Igam1997}, by the factor $\Gamma^2(l_B+1+\eta_B)/(l_B!)^2D_{l_B}(-i\eta_B)$,
where $l_B=l$.

Besides, the general equations (\ref{eq10}) and (\ref{eq13}) or the approximate
equations (\ref{eq19}) and (\ref{eq17}) are valid both for the charged case
and for the neutral one since $f_l(0)=2l+1$ and $J_l(0)=1$.
In contrast, the approximate relations (17) and (18) of Ref.~\cite{Spar2009}
are derived separately for the charged and neutral cases,
respectively, for $P_{lj}=Q_{lj}=0$.

In the two-body potential model, the ANC $C_{lj}$ for $A+a \to B$ determines
the amplitude of the tail of the $B$-nucleus
bound-state wave function in the ($A+a$) channel.
The ANC is related to the NVC $G_{lj}$ for the virtual decay $B \to A+a$
by \cite{Blokh77}
\begin{equation}
G_{lj}=-i^{l +\eta_B}\frac{\sqrt{\pi}}{\mu}C_{lj},
\label{eq20}
\end{equation}
where the combinatorial factor taking into account the nucleon identity is absorbed
in $C_{lj}$.
The numerical value of the ANC depends on the specific model used to describe
the wave functions of the $A$, $a$, and $B$ nuclei \cite{Blokh2008}.
Hence, the proportionality factor in Eq.~(\ref{eq20}), which relates NVC and ANC,
depends on the choice of nuclear model \cite{Blokh2008}.
But, as noted in Ref.~\cite{Blokh2008}, the NVC $G_{lj}$ is a more fundamental
quantity than the ANC $C_{lj}$ since the NVC is determined in a model-independent
way by Eq.~(\ref{eq11}) as the residue of the partial amplitude of the $Aa$ elastic
scattering at the pole $E=-\varepsilon $.

Using Eqs.~(\ref{eq17}) and (\ref{eq20}), one obtains
\begin{equation}
C_{lj}^2=\frac{2\kappa^{2l+1}\Gamma^2(l+1+\eta_B)}
{l!^2{\cal B}_l(\kappa,\eta_B;r_{lj},P_{lj},Q_{lj})}\,.
\label{eq21}
\end{equation}
As seen from Eq.~(\ref{eq18}),
the NVC or ANC in the effective-range approximation given by Eq.~(\ref{eq17})
or (\ref{eq21}) is expressed through the binding energy $\epsilon$
and the effective range parameters $r_{lj}$, $P_{lj}$, and $Q_{lj}$.

In the absence of Coulomb interaction ($\eta_B=0$),
expression (\ref{eq21}) can be reduced to
\begin{equation}
C_{lj}^2=\frac{2\kappa^{2l+1}}{(-1)^l(2l+1)\kappa^{2l}-
r_{lj}\kappa-4P_{lj} r_{lj}^3 \kappa^3-6Q_{lj}\kappa^5},
\label{eq22}
\end{equation}
where $r_{lj}$, $P_{lj}$, and $Q_{lj}$ are the
effective-range parameters for the purely strong interaction.
Expression (\ref{eq22}) for $l=0$ and $P_{lj}=Q_{lj}=0$
coincides with formula (3.12) of Section 3 of Chapter 3 of Ref.~\cite{BZP}
obtained for the $s$ wave.

It should be noted that expressions (\ref{eq19}), (\ref{eq17}),
(\ref{eq21}), and (\ref{eq22}) can also be applied for resonant
states of nucleus $B$ \cite{Orlov2010}. In this case,
the binding energy $\varepsilon $ should be replaced by
$-E_r+i\Gamma/2$, where $E_r$ and $\Gamma$ are the energy and width
of the resonant state of $B$ in the ($A+a$) channel, respectively.

Expressions (\ref{eq19}) and (\ref{eq21}) can be used for the analysis
of an experimental phase shift for the $Aa$ scattering at low energies
if a weakly bound state exists in the $lj$ partial wave
and if a value of the ANC is known.
In this case, a determination of the effective-range coefficients in
the sixth-order approximation (\ref{eq16}) makes it possible to reduce
the number of parameters to two.
For example, Eqs.~(\ref{eq19}) and (\ref{eq17}) can be used to express
the $r_{lj}$ and $P_{lj}$ parameters through the $a_{lj}$ and $Q_{lj}$ ones
as well as the ANC $C_{lj}$, as
\begin{equation}
r_{lj} = \frac{2\kappa^{2l}\Gamma^2(l+1+\eta_B)}{l!^2C_{lj}^2}
-\frac{4}{\kappa^2a_{lj}}+2Q_{lj}\kappa^4+
(-1)^l\kappa^{2l-1}[4J_l(\eta_B)-f_l(\eta_B)]
\label{eq23}
\end{equation}
and
\begin{equation}
r_{lj}^3 \kappa^4 P_{lj} =
-\frac{\kappa^{2l+2}\Gamma^2(l+1+\eta_B)}{l!^2C_{lj}^2}
+\frac{1}{a_{lj}}-2Q_{lj}\kappa^6+(-1)^l\kappa^{2l+1}[\half f_l(\eta_B)-J_l(\eta_B)].
 \label{eq24}
\end{equation}
By inserting Eqs.~(\ref{eq23}) and (\ref{eq24}) in the right-hand side of Eq.~(\ref{eq16}),
the truncated effective-range function $K_{lj}(k^2)$ given by Eq.~(\ref{eq16})
can be reduced to the form
\begin{eqnarray}
K_{lj}(k^2)\approx
-\frac{1}{a_{lj}}\left(\frac{k^2+\kappa^2}{\kappa^2}\right)^2+Q_{lj}k^2(k^2+\kappa^2)^2
+\frac{\kappa^{2l-2}\Gamma^2(l+1+\eta_B)}{l!^2C_{lj}^2}k^2(k^2+\kappa^2)
\nonumber \\
-(-1)^l\kappa^{2l-3}\left[\half f_l(\eta_B)(k^2+\kappa^2)-J_l(\eta_B)(k^2+2\kappa^2)
\right]k^2.
\label{eq25}
\end{eqnarray}

Equations (\ref{eq6}), (\ref{eq21}), and (\ref{eq25}) make it possible to
find $a_{lj}$ and $Q_{lj}$ if the scattering phase shift $\delta_{lj}$
and the ANC $C_{lj}$ are replaced by some experimental phase shift
at a low energy and an experimental ANC value, respectively.
The values of the $a_{lj}$ and $Q_{lj}$ parameters found in such a way
can be used for determining the $r_{lj}$ and $P_{lj}$ parameters
with Eqs.~(\ref{eq23}) and (\ref{eq24}).

\section{Applications}
\label{app}

Let us now apply these results to typical nuclear collisions.
As examples, we consider the $^{16}$O+$p$, $\alpha$+$t$,
and $\alpha $+$^3$He collisions
since ``experimental'' ANC values are known for $^{16}$O+$p\to^{17}$F
\cite{AITY2009}, $\alpha$+$t\to^7$Li \cite{Igam1997},
and $\alpha$+$^3$He$\to^7$Be \cite{Igam2010b}, respectively.
These collisions have been considered in Ref.~\cite{KB2007}
within a microscopic cluster model,
i.e.\ the generator-coordinate version of the resonating-group method,
where a direct calculation of parameters $a_{lj}$, $r_{lj}$, and $P_{lj}$
was performed at zero energy.
The exchange and spin-orbit parameters of the nucleon-nucleon effective
interaction \cite{TLT77} are fitted to the low-energy experimental phase shifts
which are fairly well reproduced.
The microscopic phase shifts and the approximate phase shifts calculated
with the effective-range expansion truncated at $k^4$
agree with each other for $E<5$ MeV.
Therefore, for a low-energy phase-shift analysis,
Eqs.~(\ref{eq6}) and (\ref{eq25}) can be safely used with $Q_{lj}=0$.

The information about the values of the ANCs from
Refs.~\cite{Igam1997,AITY2009,Igam2010b} is taken into account.
First, a single free parameter, the scattering length $a_{lj}$,
is adjusted by averaging its value obtained from Eqs.~(\ref{eq6}) and (\ref{eq25})
for several experimental phase shifts measured at low energies.
Then the $r_{lj}$ and $P_{lj}$ coefficients of the effective-range expansion
are deduced from Eqs.~(\ref{eq23}) and (\ref{eq24}).
Finally, we compare them with the coefficients obtained
with the microscopic calculation of Ref.~\cite{KB2007}.

\subsection{$^{16}$O+$p$}

\begin{figure}[ht]
\begin{center}
\epsfxsize=10.cm \centerline{\epsfbox{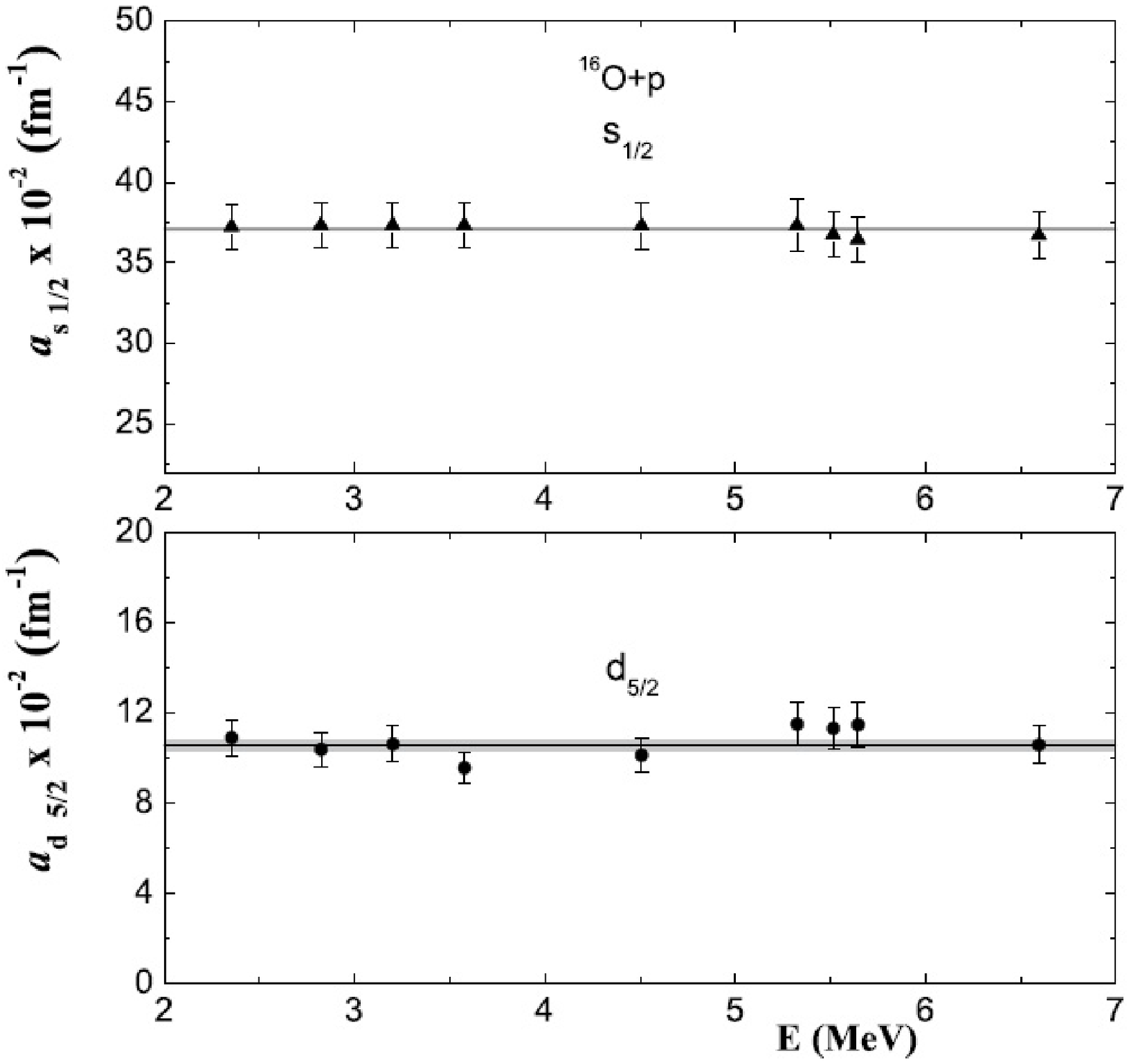}}
\caption{\label{fig11}${\rm {^{16}O}}+p$ scattering lengths for the
$s1/2$ (triangles) and $d5/2$ (circles) partial waves. 
The points are obtained with Eqs.~(\ref{eq6}) and (\ref{eq25}) 
from experimental phase shifts \cite{BHT1967} at various energies. 
The straight lines and widths of band represent our results 
for the weighted mean and its uncertainty, respectively.}
\end{center}
\end{figure}

Only the $1/2^+$ and $5/2^+$ partial waves possess bound states, at
binding energies 0.60 and 0.105 MeV, respectively. 
Hence we only analyze the experimental $1/2^+$ and $5/2^+$ phase shifts
\cite{BHT1967} corresponding to the $s$ and $d$ waves. 
The ``experimental'' ANCs for $^{16}$O+$p\to^{17}$F(g.s.; $5/2^+$) 
and $^{16}$O+$p\to^{17}$F(0.495 MeV; $1/2^+$) are known \cite{AITY2009}.
The effective-range parameters $a_{s1/2}$ and $a_{d5/2}$, and their
uncertainties, are determined from Eqs.~(\ref{eq23}) and (\ref{eq24}) 
for $Q_{lj}=0$  with the experimental phase shifts $\delta_{lj}^{\rm exp}(E)$. 
Each point plotted in Fig.~\ref{fig11} corresponds to a different energy $E$. 
The uncertainties are the averaged square errors found 
from Eqs.~(\ref{eq6}) and (\ref{eq25}), which include the experimental 
errors for the cross sections ($\sim$ 10\%) and the uncertainties 
on the ANCs.

\begin{figure}[ht]
\begin{center}
\epsfxsize=10.cm \centerline{\epsfbox{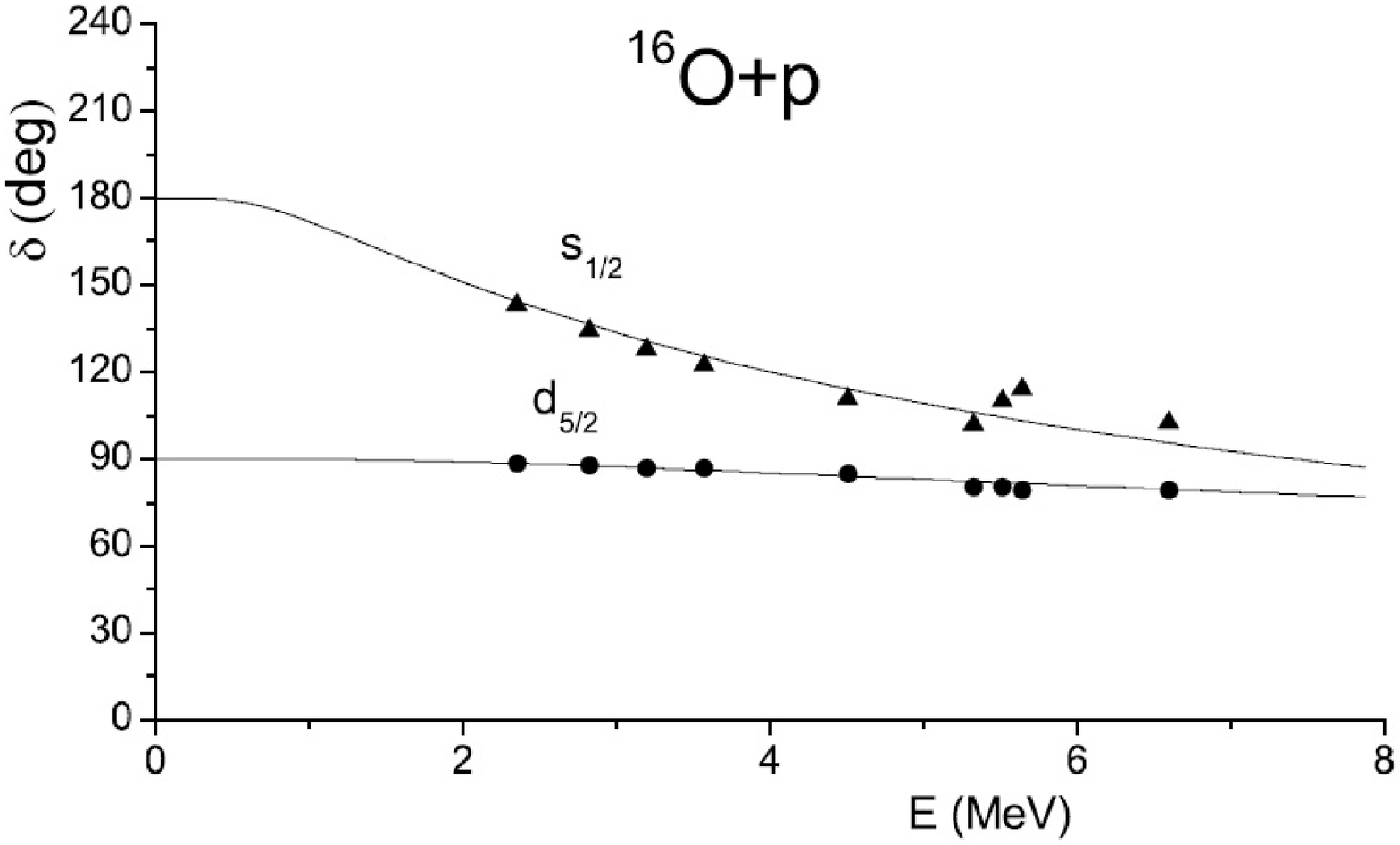}}
\caption{\label{fig1}$s1/2$ and $d5/2$ ${\rm {^{16}O}}+p$ phase shifts.
The curves display results obtained from the effective-range
expansion with parameters from Table \ref{table1}.
Triangles ($s_{1/2}$) and circles ($d_{5/2}$) represent
experimental values of Ref.~\cite{BHT1967}.
The $d_{5/2}$ phase shifts are shifted by 90 degrees for clarity.}
\end{center}
\end{figure}

The weighted means of $a_{s1/2}$ and $a_{d5/2}$ and their uncertainties
obtained from all the experimental points are given in the first and
fourth lines of Table \ref{table1}.
The corresponding mean values for the other effective-range parameters $r_{lj}$
and $P_{lj}$ obtained from the scattering lengths with Eqs.~(\ref{eq23})
and (\ref{eq24}) are also presented.
Condition (\ref{eq10}) is satisfied but it is very sensitive to the accuracy
of the effective range $r_{lj}$.
As seen from Fig.~\ref{fig1}, the phase shifts obtained with the effective-range
expansions calculated by using the parameters deduced in the present work
(solid lines in Fig.~\ref{fig1})
are in good agreement with the experimental phase shifts \cite{BHT1967}.
\begin{table}[ht]
\caption{\label{table1}Binding energy $\epsilon$ (MeV)
and effective-range coefficients for
$lj$ partial waves of various collisions:
scattering length $a_{lj}\,({\rm {fm}}^{2l+1}$),
effective range $r_{lj}$\,(${\rm {fm}}^{-2l+1}$),
and shape parameter $P_{lj}$ (fm$^{4l}$).
ANCs from the present analysis $(C^{\rm eff}_{lj})^2$ and
ANCs deduced from experiment $(C^{\rm exp}_{lj})^2$ (fm$^{-1}$).
In some lines, the ANC is obtained from the effective-range parameters
of Ref.~\cite{KB2007} (see text).}
 \begin{center}
\begin{tabular}{|c|c|c|c|c|c|c|c|c|}
\hline Collision&
$l$&$j$&$\epsilon$& $a_{lj}$       &$r_{lj}$             &$P_{lj}$     &$(C^{\rm eff}_{lj})^2$ &$(C^{\rm exp}_{lj})^2$ \\
\hline
${\rm {^{16}O}}$+$p$
&0&1/2 &0.105 &  $3708\pm48$      &$1.156\pm0.005$      &$-0.17\pm0.36$                &5700  &$5700\pm225$ \cite{AITY2009} \\
& &    &0.65  & 815 \cite{KB2007} &1.16 \cite{KB2007}   &$-0.22$ \cite{KB2007}         &53.2  &\\
& &    &0.105 & 3828              &1.16 \cite{KB2007}   &$-0.22$ \cite{KB2007}         &5850  &\\
&2&5/2 &0.60  & $1057\pm27$       &$-0.0804\pm0.007$    &$-365.6\pm161.7$              &1.09  &$1.09\pm0.11$ \cite{AITY2009}\\
\hline
$\alpha$+$t$
&1&1/2 &1.99  & $95.13\pm1.73$    &$-0.238\pm0.007$     &$39.18\pm4.90$                &9.00  &$9.00\pm0.90$ \cite{Igam1997} \\
& &    &3.75  &108.9 \cite{KB2007}&$-0.22$ \cite{KB2007}&56.99 \cite{KB2007}           &7.70  & \\
& &    &1.99  & 90.1              &$-0.22$ \cite{KB2007}&56.99 \cite{KB2007}           &5.43  & \\
&1&3/2 &2.47  & $58.10\pm0.65$    &$-0.346\pm0.005$     &$9.86\pm0.76$                 &12.74 &$12.74\pm1.10$ \cite{Igam1997}\\
& &    &4.99  &72.77 \cite{KB2007}&$-0.27$ \cite{KB2007}&26.59 \cite{KB2007}           &16.47 & \\
& &    &2.47  &69.9               &$-0.27$ \cite{KB2007}&26.59 \cite{KB2007}           &8.82  & \\
\hline
$\alpha$+${\rm {^3He}}$
&1&1/2 &1.156 & $413\pm7$         &$-0.00267\pm0.0028$  & $(2.66\pm8.39)\times10^7$    &15.9  &$15.9\pm1.1$ \cite{Igam2010b}\\
& &    &2.88  & 665 \cite{KB2007} &$-0.01$ \cite{KB2007}&$3.56\times10^6$ \cite{KB2007}&0.28  & \\
&1&3/2 &1.587 & $301\pm6$         &$-0.0170\pm0.0026$   & $(9.69\pm4.69)\times10^4$    &23.2  &$23.2\pm1.7$ \cite{Igam2010b}\\
& &    &3.99  &253.1 \cite{KB2007}&$-0.04$ \cite{KB2007}& 9212 \cite{KB2007}           &4.79  & \\
\hline
\end{tabular}
\end{center}
\end{table}

The effective-range expansion limited at fourth order obtained
with the microscopic cluster model of Ref.~\cite{KB2007} also provides
an accurate reproduction of the same experimental phase shifts
for the $s$ wave.
For a comparison, the values of the corresponding effective-range
parameters deduced in Ref.~\cite{KB2007} for the $s$ wave
are also given in Table \ref{table1}.
They are rather different although they offer an almost identical reproduction
of the experimental phase shifts.
If one uses the parameters of Ref.~\cite{KB2007} (including the energy $\epsilon=0.65$)
to compute the ANC with Eq.~(\ref{eq21}), one obtains the value 53.2 fm$^{-1}$
presented in the second last column, which differs strongly from that recommended
in Ref.~\cite{AITY2009}.
One reason for this difference is the fact that the microscopic model
provides a too large binding energy equal to 0.65 MeV.

In order to understand this discrepancy, we calculate $C_{s1/2}^2$
with the microscopic $r_{s1/2}$ and $P_{s1/2}$ but now with the exact
binding energy 0.105 MeV (see the third line in Table \ref{table1}).
The result 5850 fm$^{-1}$ is now close to the ``experimental'' value.
This is due to the good agreement for the effective range and form parameter.
Introducing this value in the pole condition (\ref{eq10}),
one obtains the scattering length 3828 fm in good agreement with 
the value in the first line of Table \ref{table1}.
However, a close reproduction of the phase shifts is important.
By fitting the exchange parameter in the effective interaction
to reproduce the experimental binding energy 0.105 MeV,
one obtains a poor reproduction of the experimental phase shifts.
The corresponding values $a_{s1/2}=4673$ fm, $r_{s1/2}=1.18$ fm, and
$P_{s1/2}=-0.23$ obtained in Ref.~\cite{KB2007} lead to
$(C_{s1/2})^2$ equal to 7142 fm$^{-1}$.
This ANC value is significantly better than the value deduced above
from the parameters of Ref.~\cite{KB2007} but it is less good
than when the phase shifts are fitted.

It is interesting to note that the effective-range expansions
lead to very similar low-energy phase shifts because the effective ranges $r_{s1/2}$
are in very good agreement.
The important difference between the scattering lengths has very little influence
because it is the inverse of this large parameter that appears in the effective-range expansion.
However, the parameters of Ref.~\cite{KB2007} allow to reliably estimate the ANC
only if the experimental binding energy is used.
\begin{figure}[ht]
\begin{center}
\epsfxsize=10.cm \centerline{\epsfbox{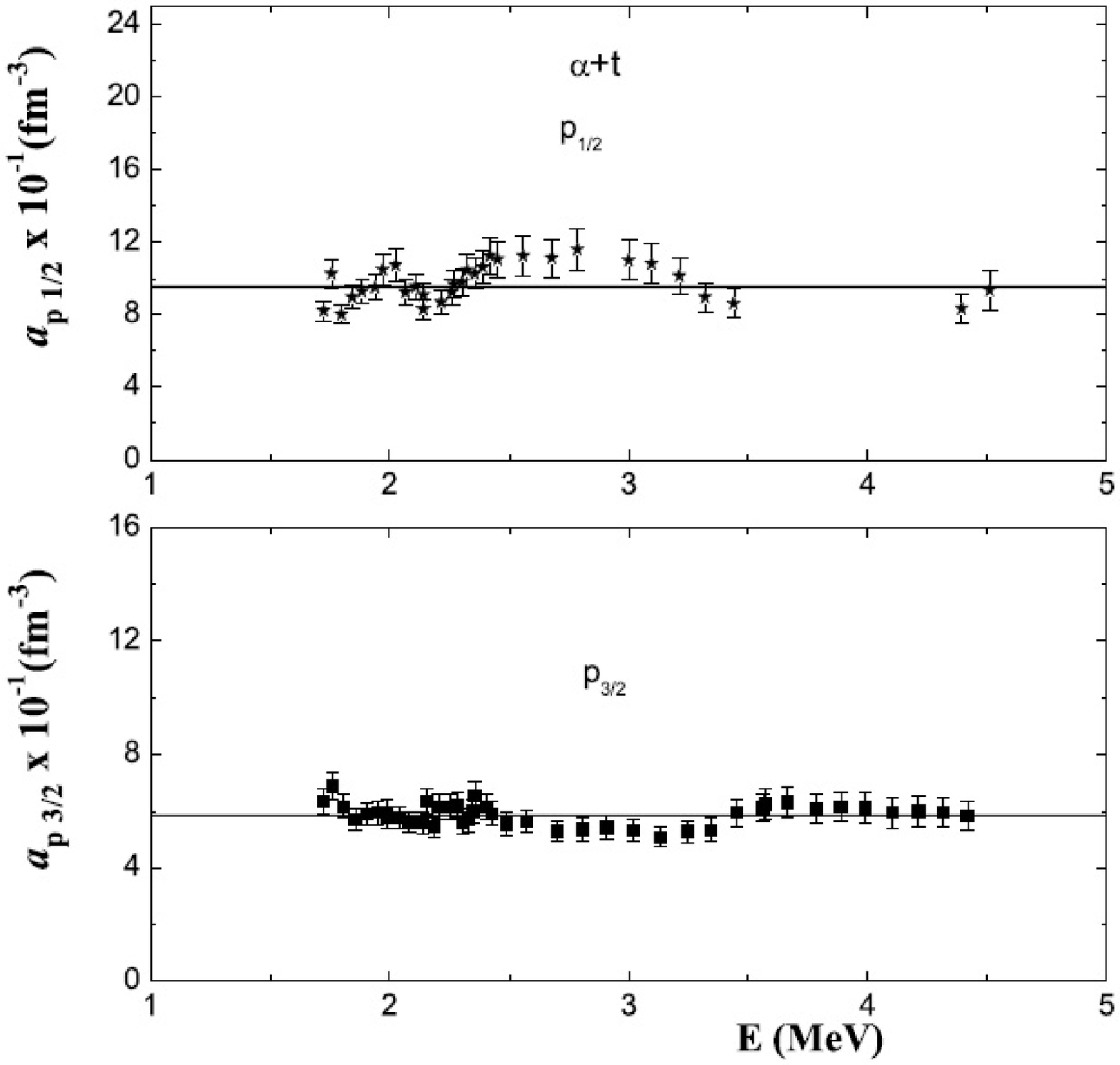}}
\caption{\label{fig22}$\alpha+t$ scattering lengths for the $p1/2$
(stars) and $p3/2$ (squares) partial waves. 
The points are obtained with Eqs.~(\ref{eq6}) and (\ref{eq25}) 
from experimental phase shifts \cite{ST1967} at various energies. 
The straight lines and widths of band represent our results 
for the weighted mean and its uncertainty, respectively.}
\end{center}
\end{figure}

Thus, for the $^{16}$O+$p$ collision, the application of formula (\ref{eq21})
together with condition (\ref{eq10}) makes it possible to
choose a set of effective-range parameters simultaneously
describing  the fundamental characteristics of the weakly bound
states of the $^{17}$F nucleus and the $^{16}$O+$p$
collision in a consistent way.

\subsection{$\alpha$+$t$ and $\alpha$+$^3$He}

\begin{figure}[t]
\begin{center}
\epsfxsize=10.cm \centerline{\epsfbox{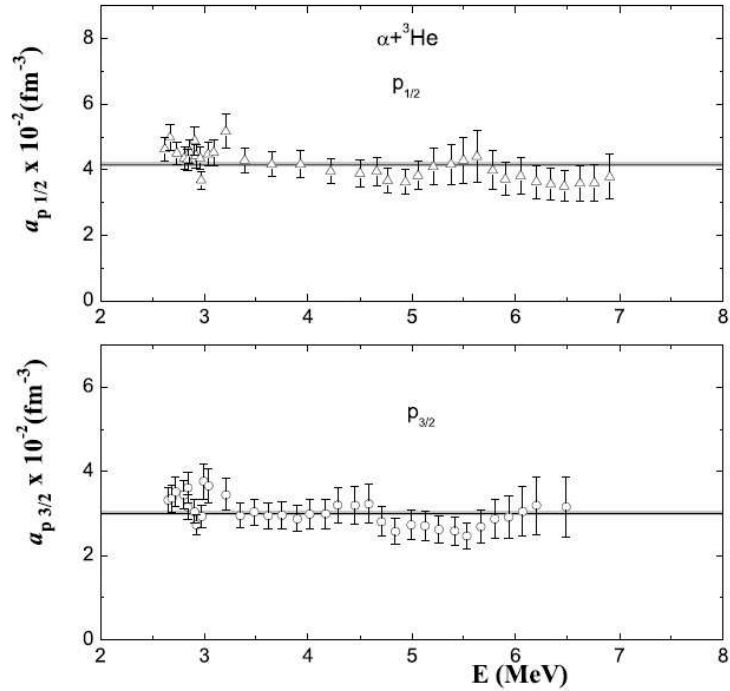}}
\caption{\label{fig33}$\alpha+^3$He scattering lengths for the
$p1/2$ (triangles) and $p3/2$ (circles) partial waves. 
See caption of Fig.~\ref{fig22}.}
\end{center}
\end{figure}

For the $\alpha$+$t$ and $\alpha$+$^3$He collisions, we only
consider the $p3/2$ and $p1/2$ partial waves which contain bound
states (see the binding energies $\epsilon$ in Table \ref{table1}).
The scattering lengths obtained as a function of experimental phase
shifts \cite{ST1967} for the $p1/2$ and $p3/2$ partial waves  of the
$\alpha$+$t$ and $\alpha$+$^3$He collisions are displayed in
Figs.~\ref{fig22} and \ref{fig33}, respectively.  The ``experimental'' ANC values
for the ground and first excited states of the $^7$Li and $^7$Be
nuclei in $\alpha+t\to^7$Li and $\alpha+^3{\rm He}\to^7$Be,
respectively, are taken from Refs.~\cite{Igam1997,Igam2010b}. The
corresponding effective-range parameters are given in Table \ref{table1}.

\begin{figure}[ht]
\begin{center}
\epsfxsize=10.cm \centerline{\epsfbox{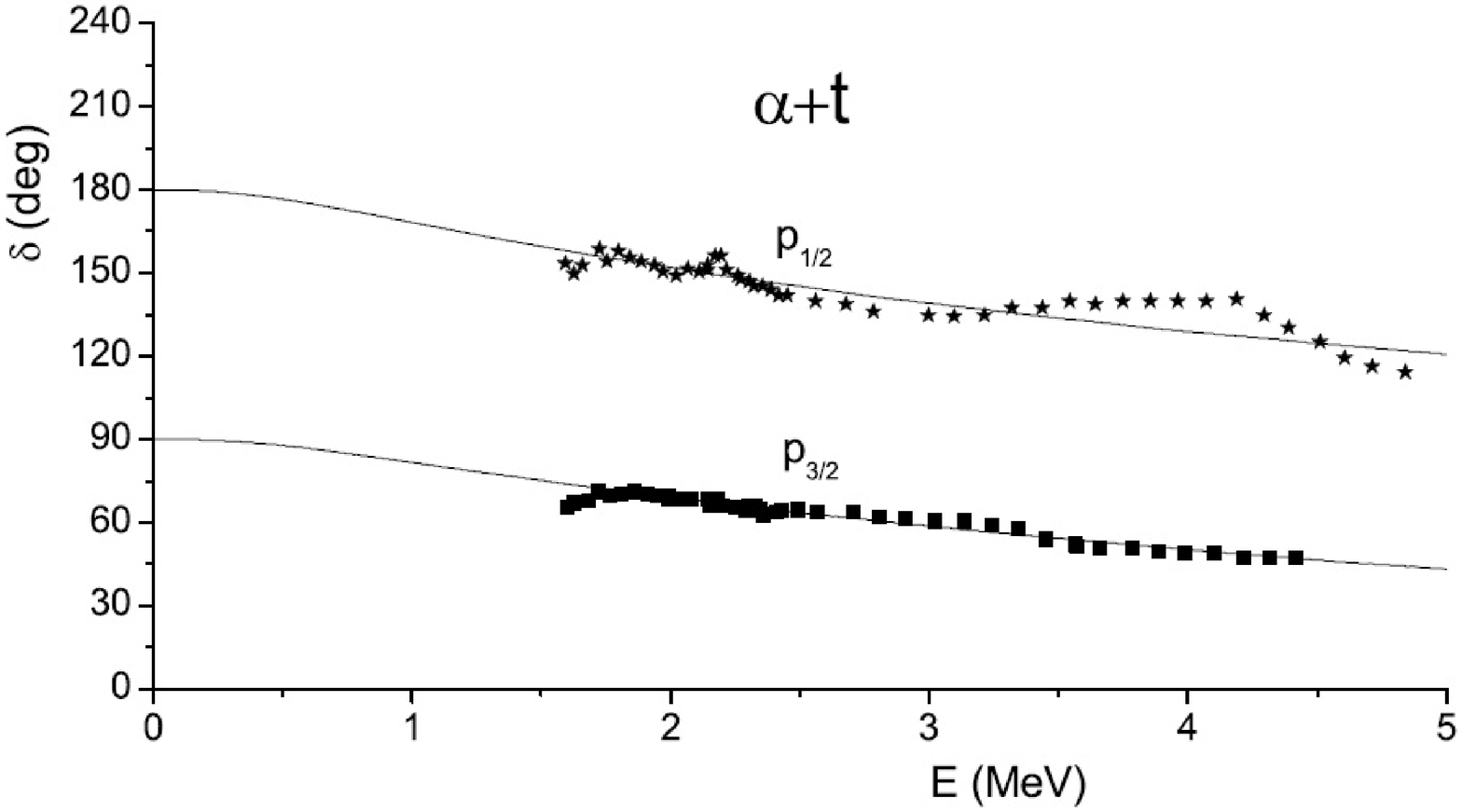}}
\caption{\label{fig2}$p1/2$ and $p3/2$ $\alpha +t$ phase shifts.
The curves display results obtained from the effective-range
expansion with parameters from Table \ref{table1}.
Stars ($p_{1/2}$) and squares ($p_{3/2}$)  represent
experimental values of Ref.~\cite{ST1967}.
The $p_{3/2}$ phase shifts are shifted by 90 degrees for clarity.}
\end{center}
\end{figure}

\begin{figure}[ht]
\begin{center}
\epsfxsize=10.cm \centerline{\epsfbox{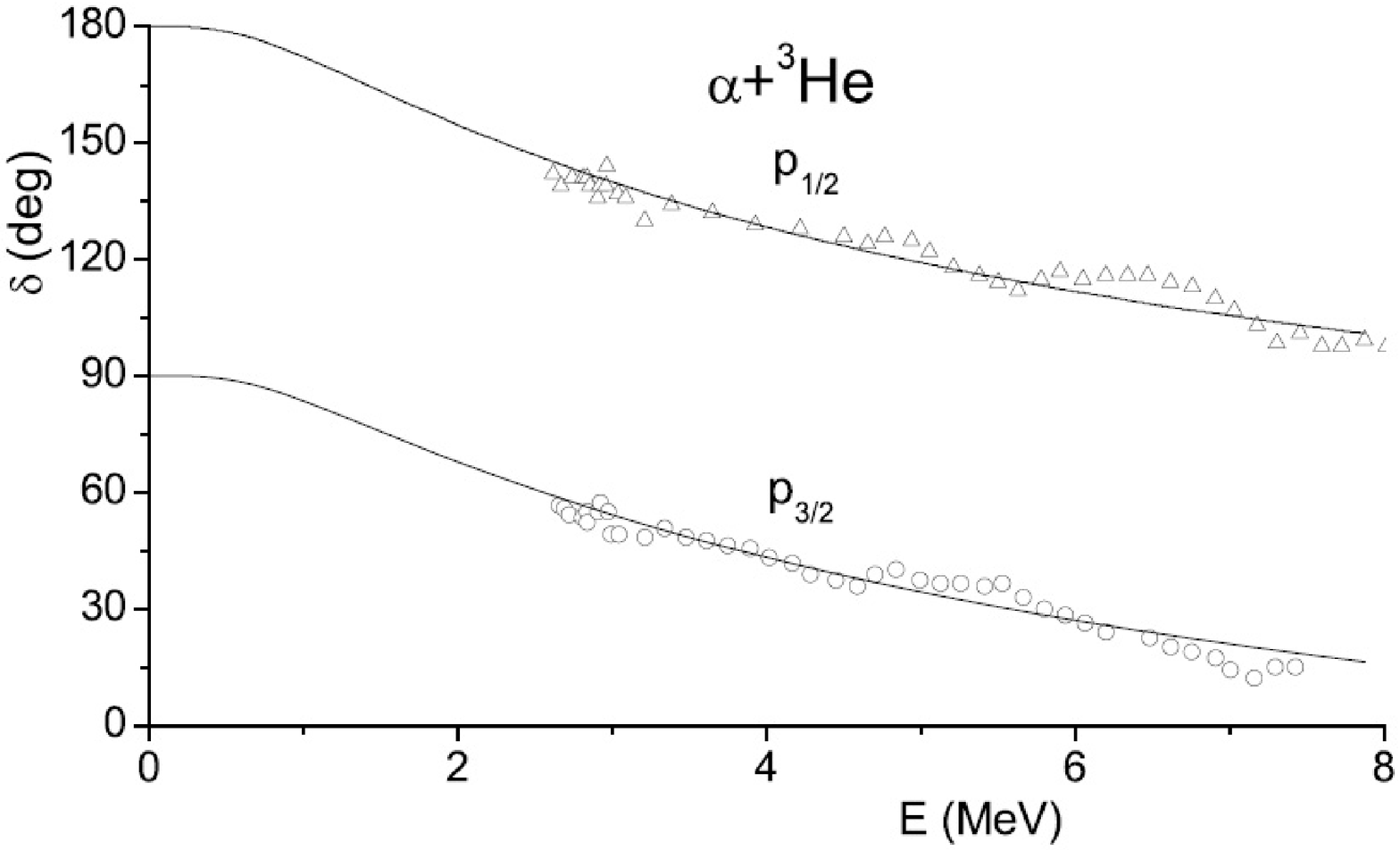}}
\caption{\label{fig3}$p1/2$ and $p3/2$ $\alpha$+$^3$He phase shifts. 
The curves display results obtained from the effective-range
expansion with parameters from Table \ref{table1}.
Triangles ($p_{1/2}$) and circles ($p_{3/2}$)  represent
experimental values of Ref.~\cite{ST1967}.
The $p_{3/2}$ phase shifts are shifted by 90 degrees for clarity.}
\end{center}
\end{figure}

As seen in Figs.~\ref{fig2} and \ref{fig3} for both collisions,
the fourth-order effective-range expansions with the coefficients
obtained in the present work
provide good parametrizations of the experimental $p_{1/2}$ and
$p_{3/2}$ phase shifts \cite{ST1967} up to at least 5 MeV.

For a comparison, the results for the effective-range parameters
obtained in  Ref.~\cite{KB2007} are presented in Table \ref{table1}.
They also reproduce the experimental phase shifts: in fact,
the curves are indistinguishable from the present ones.
But they give ANC values that differ noticeably from  ANC values
of Refs.~\cite{Igam1997,Igam2010b} (see the second last column in Table \ref{table1}),
specially in the $\alpha$+$^3$He case.
Here also, the microscopic cluster model provides too large binding
energies $\epsilon$ for both states of $^7$Li  and $^7$Be.
Hence, the values of the effective-range parameters obtained
in Ref.~\cite{KB2007} do not satisfy the bound-state energy condition (\ref{eq10}).
The  effective-range parameters obtained in the present work
and the ANC values that we use correspond to the experimental
bound-state energies and accurately verify condition (\ref{eq10}) before rounding.
\clearpage

In the $\alpha+t$ case, we replace the microscopic binding energy by the
experimental ones.
Contrary to the $^{16}$O+$p$ case, this modification does not improve
the agreement with the experimental ANC (see Table \ref{table1}).
This can be understood by the fact that the microscopic effective range
and form parameter are not very close from the present ones,
probably because the error on the binding energy is much larger.
Most likely, only a microscopic model reproducing simultaneously
the binding energy and phase shifts would overcome this drawback.
A similar problem occurs for $\alpha$+$^3$He.

\section{Conclusion}
\label{conc}

Explicit expressions for the  NVC and ANC for the virtual decay
$B\to A+a$ at an orbital momentum $l$ and the corresponding
pole condition for the energy of the bound ($A+a$) state
are derived as a function of the effective-range expansion $K_{lj}(k^2)$.
They are valid both for the charged case and for the neutral case.
These expressions are particularized for an expansion
restricted to terms up to $k^6$.
Combining these expressions with the ``experimental'', value
of the NVC (or ANC) makes it possible to reduce the number
of free parameters in the expansion to two (or to one
if the sixth-order parameter $Q_{lj}$ is chosen as zero).
This can lead to rather simple parametrizations of phase shifts
at low energies when a bound state occurs in the partial wave.

These expressions were also used for an analysis of the experimental
phase shifts for the $^{16}$O+$p$, $\alpha+t$ and $\alpha+^3$He
collisions.
The obtained coefficients of the effective-range expansions reproduce
rather well the low-energy experimental phase shifts.
These results can thus be considered as ``experimental'' values
since they fit consistently the experimental data for
both the continuum and a bound state, i.e.\
the phase shifts and the bound-state energy and ANC.

A comparison with the coefficients of the effective-range expansions
obtained in Ref.~\cite{KB2007} within a microscopic cluster model
has been performed for the same collisions.
It is useful for testing their reliability.
In spite of reproductions of phase-shifts of similar quality,
a significant disagreement is observed for the scattering length
and the deduced ANC.
It can be related to the fact that the microscopic model can not reproduce
simultaneously the phase shifts and the binding energies.
With a correct binding energy, the microscopic model allows
a good prediction of the ANC for $^{16}$O+$p$, but not for the other collisions
where the error on the energy is much larger.
This problem should be solved in future ab initio calculations
using realistic nucleon-nucleon interactions.

We think that the present results consistent with experiments both
for the continuum and for a weakly bound state should be useful to
construct nucleon-nucleus and nucleus-nucleus potentials appearing
in different nuclear models.

\acknowledgments{ R.Y. was supported in part by the Foundation for
Fundamental Research of the Uzbekistan Academy of Science (grant No
FA-F2-F077). D.B. thanks J.-M. Sparenberg for useful discussions.
This text presents research results of the Belgian Research
Initiative on exotic nuclei (BriX), program P6/23 on interuniversity
attraction poles of the Belgian Federal Science Policy Office. R.Y.
acknowledges financial support from this program. He also thanks the
Theoretical Nuclear Physics group PNTPM-ULB where this work was 
completed for the hospitality during December 2010.}

\end{document}